\pdfoutput=1
\documentclass[10pt]{article} 
\usepackage[explicit]{titlesec}
\usepackage{hyphenat}
\usepackage{ragged2e}
\usepackage{enumitem}
\RaggedRight

\usepackage[T1]{fontenc}
\usepackage{amsmath, amsthm}
\usepackage{graphicx}
\usepackage{textcomp}

\usepackage{soul}
\setul{.6pt}{.4pt}
\usepackage{geometry}
\usepackage{fancyhdr}
\usepackage[labelfont={footnotesize,bf} , textfont=footnotesize]{caption}

\titleformat{\section}
  {\normalfont}{\thesection}{1em}{\MakeUppercase{\textbf{#1}}}
\titlespacing\section{0pt}{0pt}{-10pt}
\titleformat{\subsection}
  {\normalfont}{\thesubsection}{1em}{\textit{#1}}
\titlespacing\subsection{0pt}{0pt}{-8pt}

\makeatletter
\newcommand\sixteen{\@setfontsize\sixteen{16pt}{6}}
\renewcommand{\maketitle}{\bgroup\setlength{\parindent}{0pt}
\begin{flushleft}
\vspace{-.375in}
\sixteen\bfseries \@title
\medskip
\end{flushleft}
\textit{\@author}
\egroup}
\makeatother

\usepackage[sort&compress]{natbib}
\makeatletter
\renewcommand\@biblabel[1]{\textbf{#1.}\hfill}
\makeatother

\bibpunct{}{}{,~}{s}{,}{,}
\title{Efficient Follow-Up of Exoplanet Transits Using Small Telescopes}

\author{Peter Beck, Luke Robson, Mark Gallaway, Hugh R.A. Jones, David Campbell\newline Centre of Astrophysics Research, University of Hertfordshire, College Lane, Hatfield AL10 9AB, UK}

\begin{document}

\vspace*{.01 in}
\maketitle
\vspace{.12 in}

\section*{abstract}
This paper is to introduce an online tool\footnote{http://observatory.herts.ac.uk/exotransitpredict/ [https://github.com/lukerobson/followup]} for the prediction of exoplanet transit light curves. Small telescopes can readily capture exoplanet transits under good weather conditions when the combination of a bright star and a large transiting exoplanet results in a significant depth of transit. However, in reality there are many considerations that need to be made in order to obtain useful measurements. This paper and accompanying website layout a procedure based on differential time-series photometry that has been successfully employed using 0.4m aperture telescopes to predict the expected precision for a whole light curve. This enables robust planning to decide whether the observation of a particular exoplanet transit should be attempted and in particular to be able to readily see when it should not be attempted. This may result in a significant increase in the number of transit observations captured by non-specialists. The technique and website are also appropriate for planning a variety of variable star observations where a prediction of the light curve can be made.

\vspace{.12 in}

\section{introduction}\label{introduction}
 
Several hours of continuous observing with clear skies are required to fully capture an exoplanet transit. The variety of typical weather patterns mean that opportunities to capture a transit can be infrequent at the University of Hertfordshire\textquotesingle s Bayfordbury Observatory some 35km from central London.   Many would agree with the sentiments of Julius Caesar describing British weather \textquoteleft.. the weather miserable, with frequent rain and mist\textquoteright  (Carrington 1991). Based on hourly images from the Bayfordbury All-Sky camera \sloppy (http://observatory.herts.ac.uk/allsky/) between February 2015 and February 2019, 11\% of nights were clear and 47\% of nights were mixed. Average clear time was 31\% and of this 15\% was photometric. This can be contrasted to the Paranal site in Chile where about 91\% of nights were clear based on satellite data, and 7\% were mixed (Cavazzani et al. 2011). On average 88\% of time was clear, and 70\% was photometric. (Cavazzani et al. 2015) so only a small fraction need planning with respect to arrival or disappearance of good observing weather. Cavazzani et al (2011) only uses 3 data points per night rather than the hourly measurements for Bayfordbury (although there is likely to be less variable weather than Bayfordbury) and only looks at the clarity of the atmosphere (i.e. no criteria for the moon). Thus, at a site with a relative lack of clear skies, like Bayfordbury, careful planning is required in order to deploy precious observational resources efficiently. In particular, it is imperative to only attempt to capture an exoplanet transit which will take several hours when there is a reasonable expectation of photometric conditions.

The problem is essentially deciding under photometric conditions if one can adequately capture a predicted transit with a star of a given catalogue magnitude, a given transit depth for the predicted range of air mass at the observatory\textquotesingle s location and still have an adequate image cadence whilst still maintaining detector linearity. Only when a transit meets these criteria should an attempt be made to use prime telescope time to capture a transit.
Since the seminal first detection of an exoplanet HD 209458b by photometry (Charbonneau et al. 2000; Henry et al. 2000), the search and characterisation of exoplanets has become a major focus of interest for precision photometry and in particular for many ground-based projects such as WASP (Pollacco et al. 2006) and HAT (Vanhuysse et al. 2011), and in turn to space missions such as COROT (Deeg 2013), Kepler (Batalha et al. 2010) and TESS (Ricker et al. 2009).

For an experienced observer, capturing a transit of exoplanet HD 209458b is technically relatively easy with the Bayfordbury Observatory 40cm aperture Meade LX200 telescopes equipped with CCD cameras due to its relatively high catalogue magnitude (V=7.63mag) and large depth of transit (0.0162mag). However, inspection of transit predictions with other stars such as those readily available online via the Exoplanet Transit Database website of the Czech Astronomical Society (Poddany et al. 2010) or Jensen (2013) indicate that the targets have a wide range of magnitudes and depths of transit. For example, WASP 52b has a host star magnitude of V=12.0mag and a transit depth of 0.0296mag. It is not immediately obvious that this transit could be captured using the Bayfordbury Observatory telescopes, but transits have now been successfully observed many times. That said, capturing these transits required careful planning and the use of precision photometry. 

This paper draws on the methodology of Beck (2018) that has been used to predict the expected precision for a target (and associated reference stars) given their catalogue magnitudes, the user selected exposure time and expected range of values for air mass. Using a formulation tailored to their equipment, other observatories with modern cameras should be able to effectively plan their observations and conduct research in areas where they might not have realised the potential capability of their equipment. An online software tool has been developed to predict viable observing opportunities for specific exoplanet transits. This includes a suitable allowance for capturing the light curve both before and after the transits. The key function of the tool is to simulate the transit including appropriate errors to give a predicted dataset and graphical display.

\section{Methods}
\subsection{System Model}\label{System Model}

In order to predict whether an exoplanet transit can be satisfactorily captured, one first needs to be able to predict the errors on an observation from all sources. Many of the sources contributing errors such as target noise are common to all observatories/ telescope/ camera combinations. Some embedded terms such as the observatory\textquotesingle s altitude are specific to a given observatory. Each telescope/ camera combination will have specific design characteristics and consequently will have unique calibration values. The calibration values should be updated over time as calibrations such as flat field will change. The precision also depends on the observations being made and in particular the chosen target and reference stars, the selected exposure time and the values of air mass during the observations. For example, long exposures needed for satisfactorily capturing faint targets means fewer measurements and less well defined light curves. On the other hand, bright targets can result in a very high number of data points, with great care needed to not exceed the pixel linearity limit, especially if the value of air mass changes significantly during an observation.
The total Standard Deviation (SD) for the target star and reference star is needed and we present a condensed version of Beck (2018) and in the frequently asked questions (FAQ) of the online tool. The predicted precision on the target alone in ADU units is represented by the following equation from Southworth et al. (2009)
\begin{equation}
    (\sigma_{\textrm{total}})_{\textrm{target}}=\sqrt{(\sigma_{\textrm{bias}}^2+\sigma_{\textrm{dark}}^2+(\sigma_{\textrm{flat}})_{\textrm{target}}^2+(I_{\textrm{RMS}} )_{\textrm{target}}^2+\sigma_{\textrm{sky}}^2+\sigma〗_{\textrm{target}}^2 )}
\end{equation}

where $\sigma_{\textrm{bias}}$ is the bias noise and is a constant for a given camera. Its value is established from calibration measurements and is defined by equation (2), $\sigma_{\textrm{dark}}$ is the dark current noise for the CCD camera and increases linearly with the user defined exposure time ($t_{\textrm{exp}}$) as defined by equation (3), $\sigma_{\textrm{flat}}$  is the flat field noise for the CCD camera obtained from dome calibrations and is defined by equation (4) to (5), $I_{\textrm{RMS}}$ is scintillation noise from air turbulence as defined by equations (6) to (7), $\sigma_{\textrm{sky}}$ is sky noise at a specific observation location is defined by equation (8) and $\sigma_{\textrm{target}}$ is the target noise defined by equation (9).

The bias and dark biases for a particular camera are established from calibrations and represent the variation in each pixel in units of ADU as outlined in Beck (2018), where $t_{\textrm{exp}}$ is the user selected exposure time and $\dot{\sigma}_{\textrm{dark}}$ is the value of the dark gradient of a plot of exposure time vs dark, 

\begin{equation}
\sigma_{\textrm{bias}}=\textrm{constant}
\end{equation}

\begin{equation}
\sigma_{\textrm{dark}} = \dot{\sigma}_{\textrm{dark}} * t_{\textrm{exp}}
\end{equation}

The flat field for the particular CCD camera, the scintillation noise and the target error depend on the target count $(N_{\textrm{target}})$. For the flat field a count of 1 gives the flat field noise for $N_{\textrm{target}}=1$, where $\sigma_{\textrm{flat}}$ is the standard deviation per pixel averaged over the flat images, $(N_{\textrm{target}})_{\textrm{flat}}$  is the total ADU count per pixel averaged over the flat images, 
\begin{equation}	
(\sigma_{\textrm{flat}})_{(N_{\textrm{target=1}})} = \sigma_{\textrm{flat}}/(N_{\textrm{target}})_{\textrm{flat}}
\end{equation}

Thus the general case for the flat field noise with any count
$(N_{\textrm{target}})$ is
\begin{equation}
(\sigma_{\textrm{flat}})_{\textrm{target}} =(\sigma_{\textrm{flat}})_{(N_{\textrm{target=1}})} N_{\textrm{target}}		
\end{equation}

Atmospheric extinction and scintillation can be one of the largest sources of error in photometric observations (Balona, Janes, \& Upgren 1995), (Gilliland \& Brown 1992), (Hartman, et al. 2005), (Ryan \& Sandler 1998). The scintillation error with a small telescope and a bright target (Castellano et al. 2004) is shown to be the dominant source of error. It is important to maximise the exposure time since the longer exposure time, the lower noise due to scintillation (Dravins et al. 1998; Young, Dukes, \& Adelman 1992) since the normalised scintillation noise is given by

\begin{equation}
    \sigma_{\textrm{scint}}=0.004D^{(-2/3)} X^{(7/4)} e^{(-h/H)} (2t_{\textrm{exp}})^{-0.5} 
\end{equation}

where D is the telescope aperture (m), X is the air mass, h is the altitude of the telescope (m), H=8000 (m) is the scale height of the atmosphere.
$(I_{RMS})_{\textrm{target}}$ is the scintillation noise from air turbulence for the target obtained by scaling $\sigma_{\textrm{scint}}$ for the total ADU count for the target
\begin{equation}
    (I_{\textrm{RMS}})_{\textrm{target}}=N_{\textrm{target}} \sigma_{\textrm{scint}}	
\end{equation}

The sky SD $(\sigma_{\textrm{sky}})$ is established from values obtained from an analysis of areas of sky in several images in the absence of a target. It is recognised that from this equation it will give a ‘typical’ sky condition of the images that are analysed to give
\begin{equation}
    \sigma_{\textrm{sky}} =\sqrt{\sigma_{\textrm{sky}}^{'2} - \sigma_{\textrm{cal}}^{2}-(I_{\textrm{RMS}})_{\textrm{sky}}^{2}}    
\end{equation}

where $\sigma_{\textrm{cal}}^{2}= \sigma_{\textrm{bias}}^{2} + \sigma_{\textrm{dark}}^{2} +  \sigma_{(\textrm{flat(sky)})}^{2}$ and $(I_{\textrm{RMS}})_{\textrm{sky}}^{2} =(N_{\textrm{{sky}}} \sigma_{\textrm{scint}})^{2}$,   where $N_{\textrm{sky}}=\textrm{constant}* t_{\textrm{exp}}$  and where the constant is the ADU count of the sky background for 1 second, $\sigma_{\textrm{sky}}$ is the background noise taken from a region of sky essentially devoid of stars.
The standard assumption is made that the target signal follows Poisson noise characteristics as the counts are at a constant rate with probability of independent of time since last event and consequently the target noise is given by 
\begin{equation}
\sigma_{\textrm{target}} = \sqrt{N_{\textrm{target}} }
\end{equation}
The derivation of a predicted value for $N_{\textrm{target}}$ has been condensed down to a series of equations that after working out the constants for the equipment just need the catalogue magnitude of the target star, exposure time and the air mass. The key relationship between catalogue magnitude and instrument magnitude, for an air mass of 1.0 and an exposure time of 1.0s for a particular telescope/ camera combination is obtained from a series of observations. 
Many historical observations are needed that cover a wide range of target magnitudes, exposure times, values of air mass and observing conditions. The only pre-requisite in the choice is that there are no clear problems with the image and that there were no obvious filling of a pixel well to beyond its linearity limit. 

\begin{equation}
[m'_{\textrm{t=1s,X=1.0}}]_{\textrm{predicted}} = Gradient* m_{\textrm{target}} + Bias
\end{equation}
		     
Where Gradient and Bias come from the trendline of the Catalogue to Instrument Magnitude plot, $m_{\textrm{target}}$ is the catalogue magnitude of the target,
$[m'_{\textrm{t=1s,X=1.0,target}}]_{\textrm{predicted}}$ is the predicted instrument magnitude for the target with an exposure time of 1.0s and airmass of 1.0.
\begin{equation}
   [m'_{\textrm{t=1s,X=X}} ]_{\textrm{predicted}} = [m_{\textrm{t=1s,X=1.0}}^{'} ]_{\textrm{predicted}} - \varepsilon(1.0-X)	 
\end{equation}
where $\varepsilon$ is the airmass extinction coefficient. $[m_{\textrm{t=1s,X=X}}^{'} ]_{\textrm{predicted}}$ is the predicted instrument magnitude for the target with an exposure time of 1.0s and the airmass is the observed airmass.
\begin{equation}
[(N_{\textrm{target}})_{\textrm{(t=$t_{\textrm{exp}}$ X=X)}}]_{\textrm{predicted}}= t_{\textrm{exp}}  10^{-[m_{{(\textrm{t=1s,X=X})}}^{'} ]_{\textrm{predicted}}/2.5} 
\end{equation}

where $[(N_{\textrm{target}})_{\textrm{(t=$t_{\textrm{exp}}$ X=X)}}]_{\textrm{predicted}}$ is the predicted ADU total count for the target with an exposure time of texp and airmass of X. The best fit line gives the equation that can then be used to predict the value of instrument magnitude for the target for t=1.0s and X=1.0 based solely on the target\textquotesingle s catalogue magnitude.

\begin{figure}[t] 
\centering
\includegraphics[ trim={2.0cm 2.5cm 2.2cm 2.5cm},clip, width=0.9\textwidth]{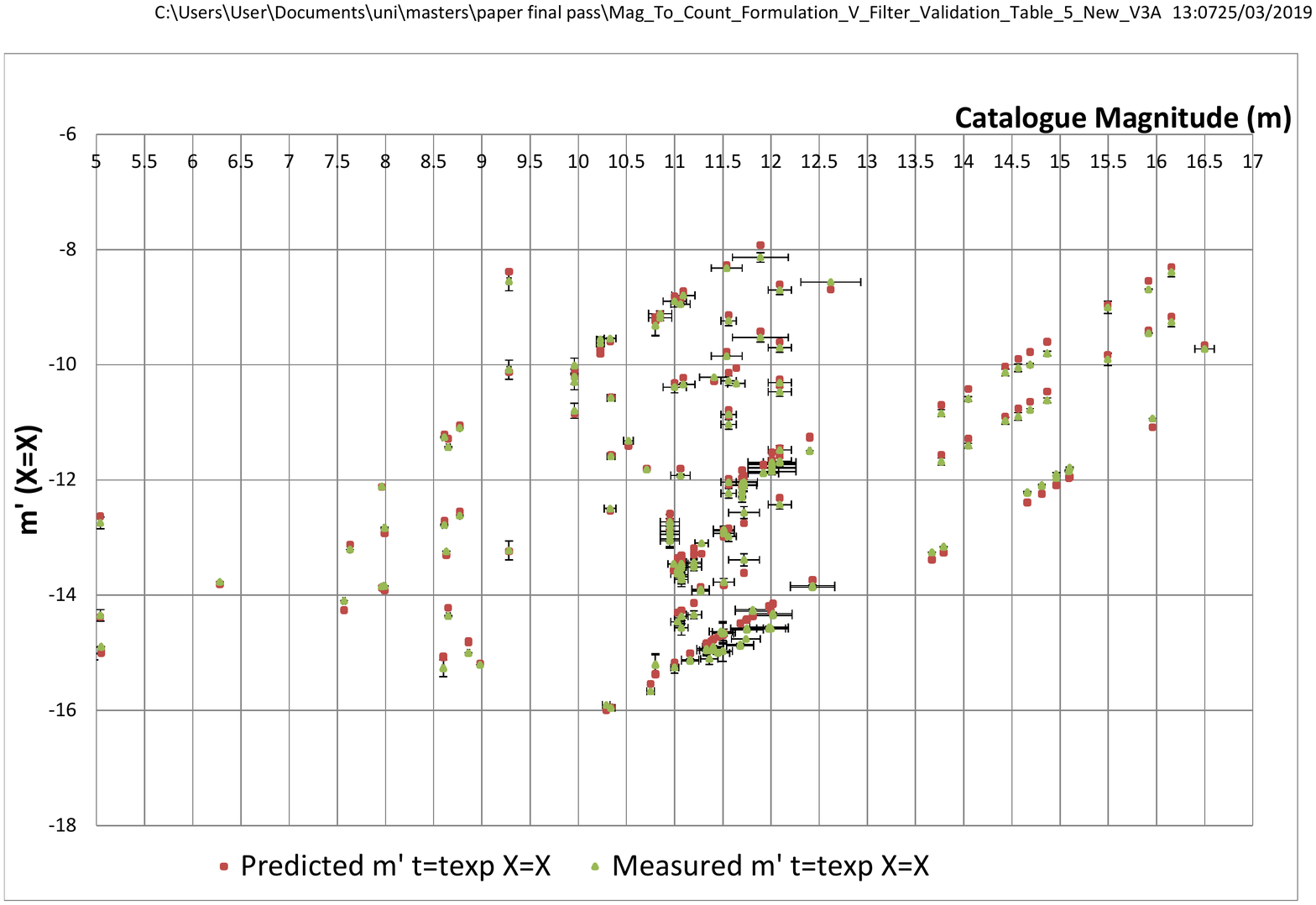}
\caption{ Predicted and measured
$(m'_{\textrm{t=$t_{\textrm{exp}}$ X=X}})$ versus catalogue magnitude (m)
}
\label{fig1}
\end{figure}

Having derived the best fit line (eq.10), the formulation was then verified by using the catalogue magnitude to back calculate a predicted value for instrument magnitude for each image and then subtract the predicted instrument magnitude from the measured instrument magnitude to give $(\Delta m^{'})$. The average and standard deviation of the values of $\Delta m^{'}$ gave an estimate of the intrinsic error in the estimate of 0.00045mag and 0.195mag respectively.

It was considered essential to establish how robust equation (10) was since the derivation of  $[m'_{\textrm{t=1s,X=1.0,target}}]_{\textrm{predicted}}$ was considered to be fundamental. Consequently a significant exercise was undertaken to take an even larger sample of 211 completely independent observations from the Bayfordbury Observatory archive of images with different targets covering a wide range of catalogue magnitude, exposure time and implicitly air mass and observing conditions to validate the range of conditions that the formulation would hold. Again the catalogue magnitude was used to back calculate a predicted value for instrument magnitude for each image and then this value was compared with the measured instrument magnitude to give $(\Delta m^{'})$. 

This validation process produced Figure 1 where a matching set of predicted instrument magnitudes (in green) and measured instrument magnitudes (in red) were produced from a large number of archived images. Inspection of Figure 1 shows that a good match was achieved over a range of catalogue magnitudes, exposure times and air masses.

\subsection{Comparison of Results From Different Telescopes}

Section \ref{System Model} identified a series of equations that have been formulated to predict the precision for each measurement of instrument magnitude from a telescope whose performance has been characterised. Light curves have been produced from the Chris Kitchen Telescope (CKT), James Hough Telescope (JHT) and Robert Priddey Telescope (RPT) robotically controlled telescopes at the Bayfordbury Observatory. The CKT configuration had been used to quantify the parameters in the SD equations and is equipped with an SBIG STL-6303E CCD camera and a Meade LX200 series telescope. The JHT also has a Meade LX200 series telescope and has the same camera designation as the CKT and consequently the SD parameters should be close to those for the CKT. However, whilst the RPT telescope also has a Meade LX200 series telescope, it has a significantly different scientific camera (Moravian Instruments G4-9000). Consequently any tolerances derived for the RPT using the CKT equations will not be fully representative as several of the telescope specific parameters have not been derived for the RPT. The field of view of the CKT and JHT telescopes being 23.4\textquotesingle x15.6\textquotesingle whereas that for the RPT is noticeably wider at 31.2\textquotesingle x31.2\textquotesingle . This implies that the FWHM for a star with the RPT occupies a relatively smaller area on the chip compared to that with same field of view as the CKT/JHT telescopes and consequently would be expected to reach a linearity limit sooner.

The CKT, JHT and RPT robotically controlled telescopes were used to independently simultaneously capture the light curve of a transit of the deeply eclipsing cataclysmic variable UX Uma on the 2017 June 1st to demonstrate consistency with measurements taken by different telescopes.
\begin{figure}[ht] 
\centering
\includegraphics[ trim={1.9cm 2.5cm 2.5cm 2.5cm},clip, width=0.9\textwidth]{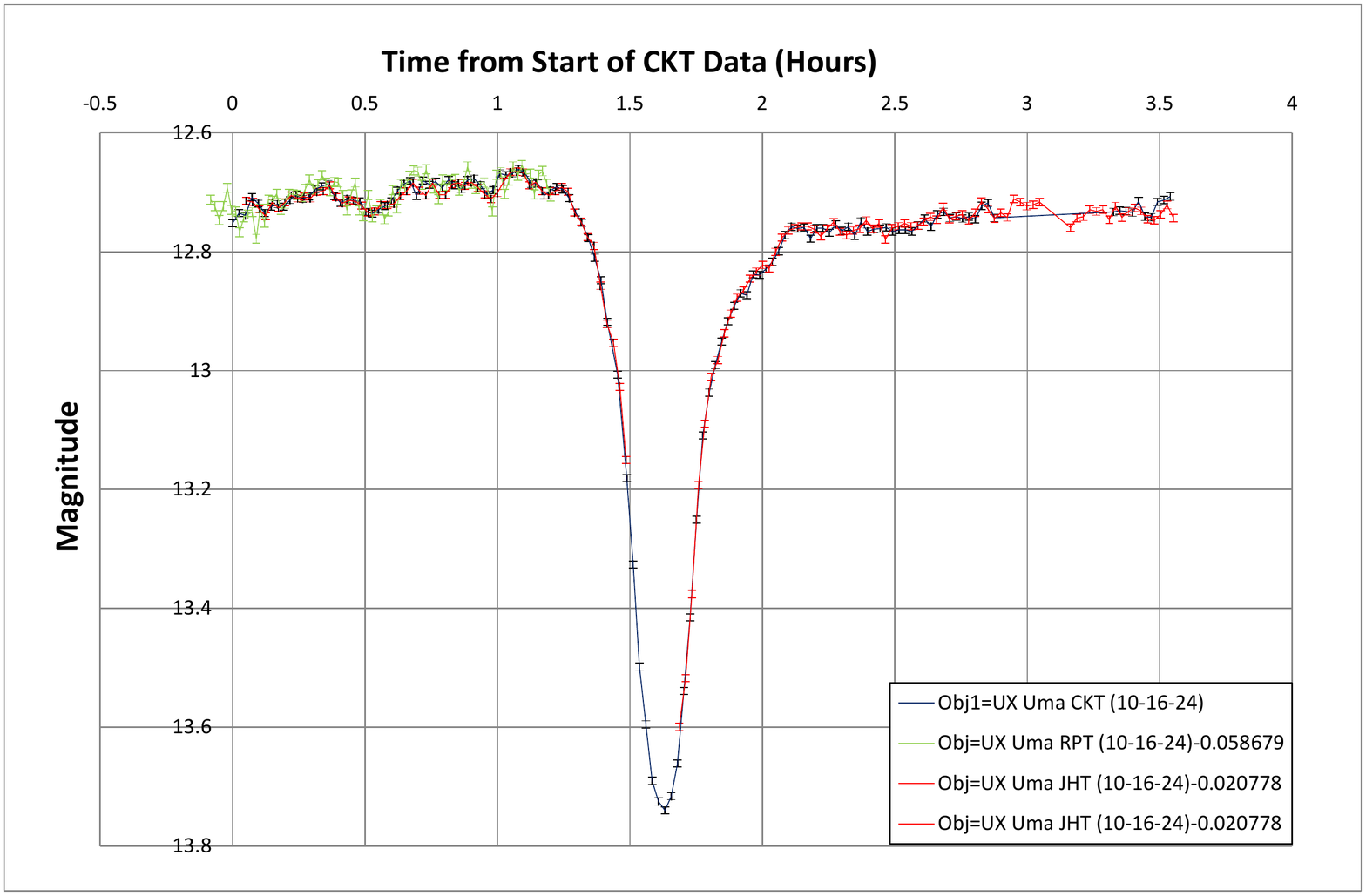}
\caption{ UX UMa eclipse of 2017 June 1st taken using CKT, JHT and RPT telescopes.
}
\label{fig2}
\end{figure}
Nominally, the light curves from the three telescopes should be identical as the equipment is similar and the observations have been conducted at the same time. However, tolerances will mean that individual ADU values will be different. The objective of this exercise was to overlay the three light curves and verify that their tolerance values typically encompass the other light curves. Achieving such a match verifies that the calculated tolerances are consistent with the observations. By using the results from different telescopes, and knowing the tolerances on individual values offers the prospect of weighting each data point according to its SD value and in turn generate a composite plot. Consequently the eclipse of UX UMa could have a light curve with significantly more data points than from a single telescope, plus the tolerances will enable the user to determine if a variation in the light curve is a plotted light curves from the different telescopes. 

Despite the significant variation in magnitude during the eclipse, Figure 2 shows that the CKT and JHT light curves and associated tolerances display a good degree of correlation. In the case of the JHT telescope, three anomalous data points have been removed that occurred shortly after the telescope conducted a re-focus and consequently the JHT light curve is drawn with two separate lines. Consequently it is concluded that the predicted precision is consistent with the observed precision. Further, simultaneous observations of the same star would warrant plotting the interpolated differences between the observations, with the width of the distribution compared to the predicted uncertainty for several examples though the opportunities to have multiple telescopes on one target are limited. 
\subsection{Transit Follow-up Tool}\label{Transit Follow-up Tool}

The transit follow-up tool includes the equations defined by the system model (Section \ref{System Model}) and provides the recommended observing times for known and yet to be confirmed exoplanet transits like targets of interest from TESS. Although predicted transits are available from web sites such as the Exoplanet Transit Database website of the Czech Astronomical Society (Poddany et al,. 2010) and Jensen (2013), the recommended transits from the Transit Follow-up Tool exclude any that cannot be satisfactorily observed in their entirety and in particular do not meet one or more of the following constraints.
\begin{enumerate}[label=\alph*)]
\item	Observing must be in the period between the scheduled start and end of robotic observing (sun below the horizon by at least 12 degrees of the horizon).
\item	Observing must include the user specified duration to capture the images both before and after the end of a transit.
\item	Must not exceed the allowed telescope elevation angle.
\item	Must still have sufficient predicted precision after allowing for variation in air mass during the observing period.
\item	The predicted precision must be sufficiently small for a specified transit depth.
\end{enumerate}

Consequently this sub-set of predicted transits identifies a limited number of key occasions when the transits are particularly suitable for observing. 

The main programming for the online transit follow-up tool is done in PHP and JS and as a website also HTML. The source code and documentation is included at the software repository GitHub (https://github.com/lukerobson/followup). Beyond providing a variety of tools to plan future exoplanet transit observations the software also produces RTML code that telescope software can use to operate a chosen telescope automatically. Below we illustrate some of the prediction and plotting tools the software provides. The data for potential exoplanet transit events for the online tool comes from the NASA exoplanet archive (2019) as an aggregator of exoplanets both confirmed and tentative whilst data for TESS targets of interest comes separately from ExoFOP (2019). 

In Figure 3, we show two examples of the use of the tool for observed and simulated observations for the transits of WASP-10b and WASP-52b with the online tool providing a number of other examples. The predicted observation is based on the known transit depth, airmass, length of transit and exposure time, including the determination of the standard deviation for each simulated point in the light curve using the air mass during the observations. The observed error bars are derived following the data reduction of the recorded images using AstroimageJ (Collins et al. 2017) as described in the online tool.

\begin{figure}[ht] 
\centering
\includegraphics[width=0.9\textwidth]{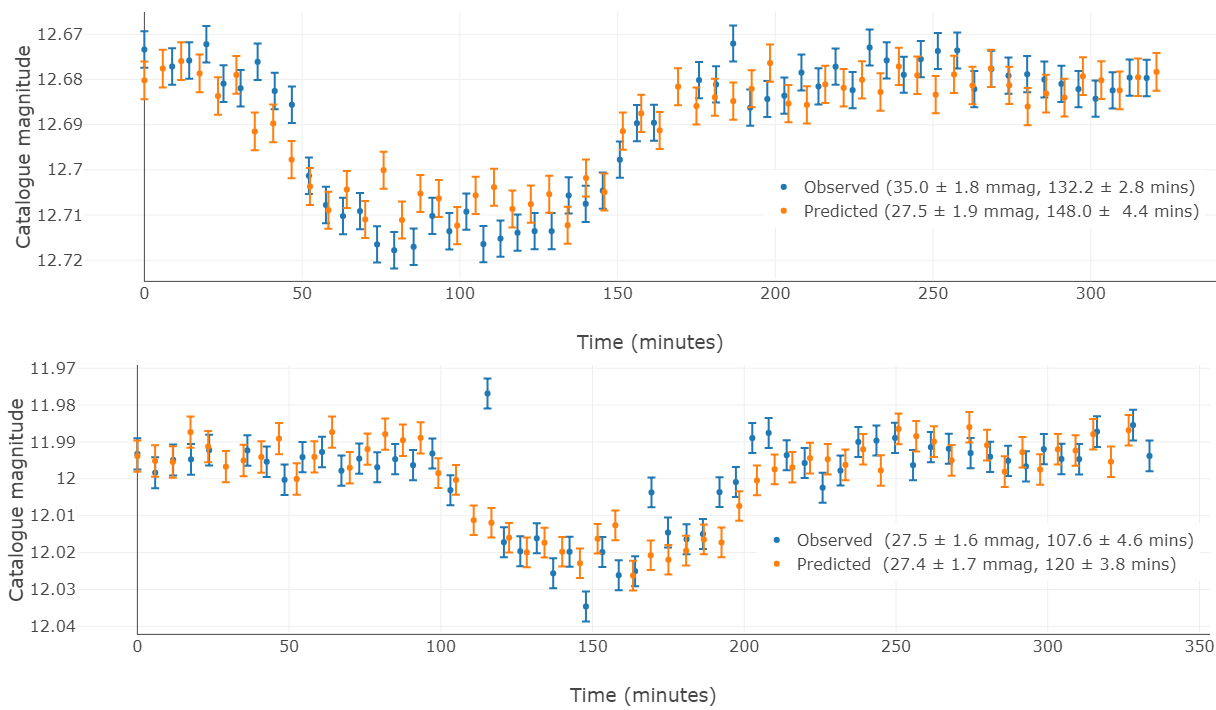}
\caption{ Plots of Observed and Predicted Light Curve from the online tool. The top plot is of WASP-10b (for a $V_{Johnson}$-band transit 2016 September 13th), the bottom plot is WASP-52b (for a $V_{Johnson}$-band transit 2016 September 23rd). The blue series shows the observed data determined using AstroimageJ and the orange series is the predicted data based on a read-out time of 50s.}
\label{fig3}
\end{figure}
\begin{figure}[b] 
\centering
\includegraphics[trim={0 0.1cm 0 0.5cm},clip,width=0.88\textwidth]{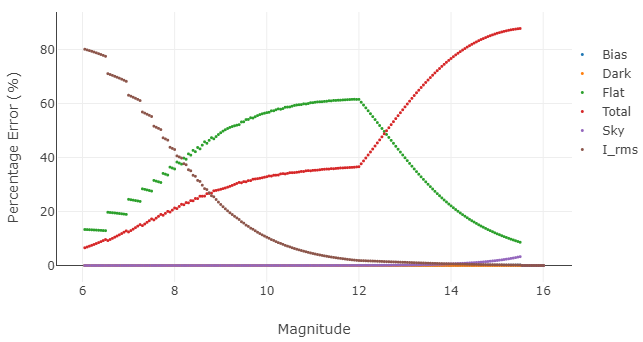}
\caption{ Breakdown of the percentage error with the CKT as a function of magnitude.}
\label{fig4}
\end{figure}
The observed light curves appear to be reasonably matched by the simulated data and are consistent at approximately two sigma as measured by the online tool provided by (Poddany et al. 2010). The predicted plot is based on the reference values as imported from the NASA exoplanet archive. It should be noted that the predicted values for depth and duration are sensitive to the precise details of the chosen sampling. The simulated data assume a constant detector read-out time but for a variety of reasons the observed data is rarely so uniform. It is also likely that some of the difference between the observed and predicted curves are intrinsic to the stars. For example, various authors have shown evidence for stellar activity modifying observed transits in terms of both the depth and duration of the transit. 
\begin{figure}[t!] 
\centering
\includegraphics[width=0.9\textwidth]{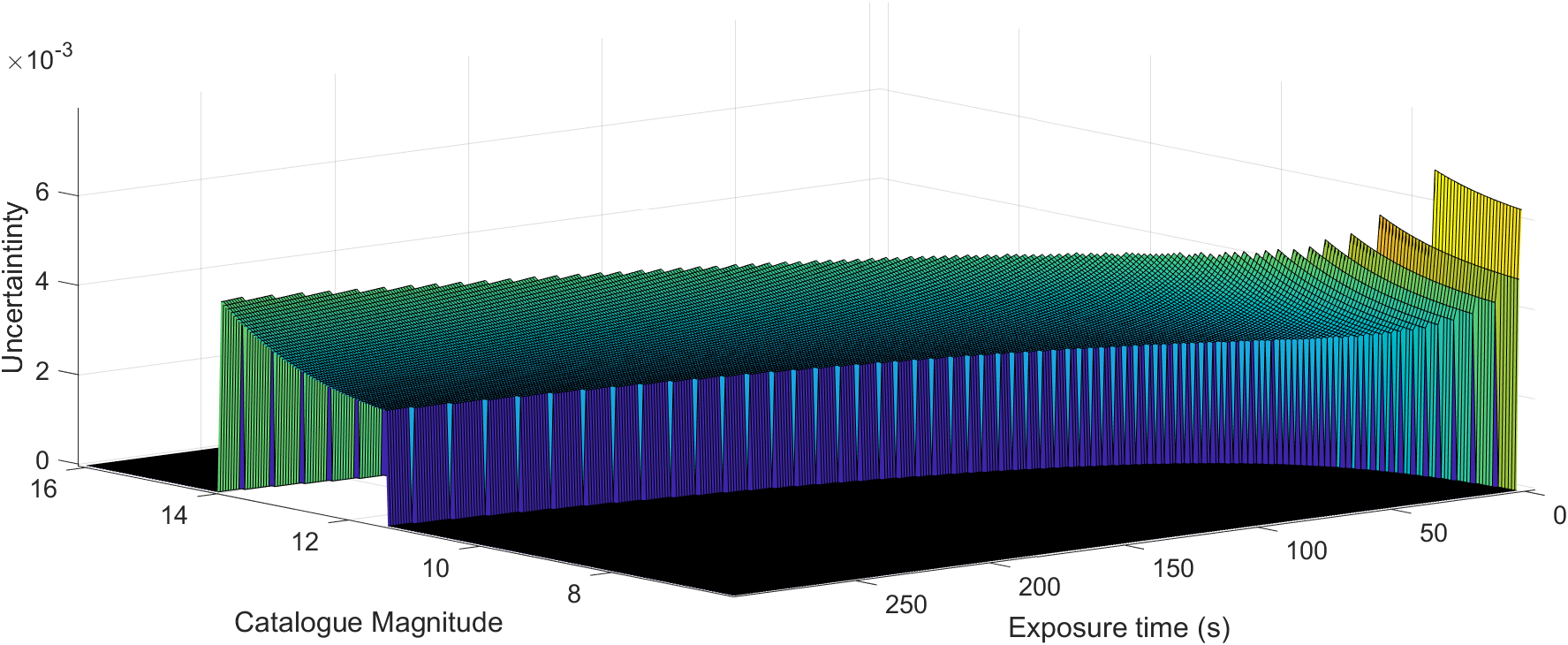}
\centering
\includegraphics[trim={3cm 0 3.5cm 1cm},clip,width=0.9\textwidth]{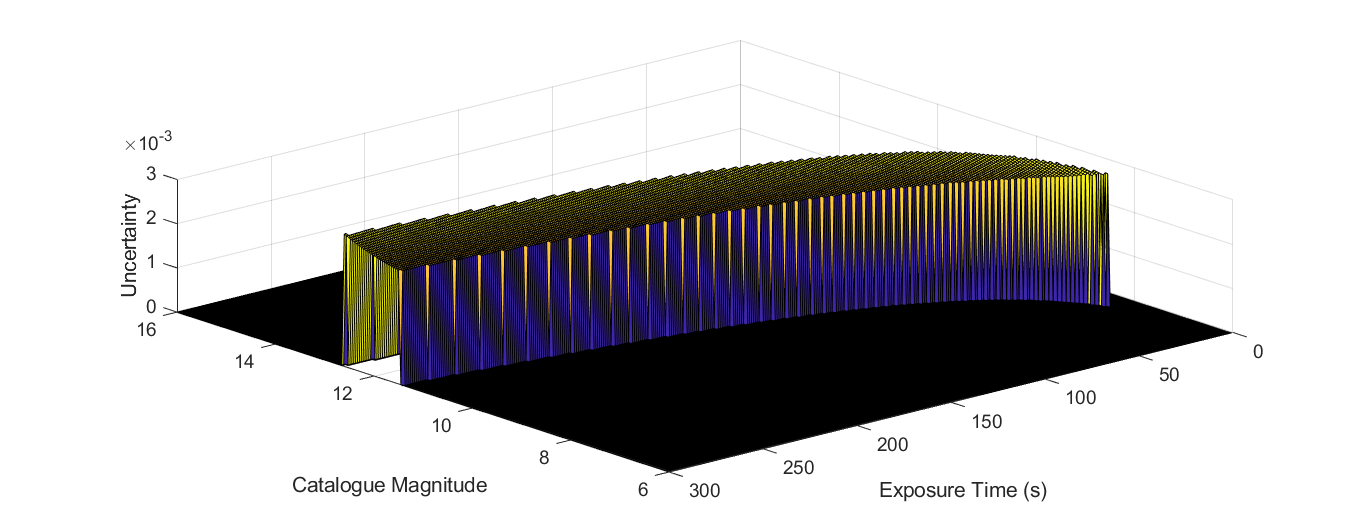}

\caption{ Predicted performance of the CKT over a range of magnitudes and exposure times (top) and over all precisions (bottom) with a limited precision \textless 3mmag.}
\label{fig5b}
\end{figure}
The scenario is that the transiting planet can be occulted by starspots leading to a modulation of transit depth with stellar rotation phase, e.g., Croll, Rappaport \& Levine (2015). For example, Maciejewski et al. (2011) identify a worst case of a 3mmag variation in WASP 10 in a 4 hour period due to spots compared to the transit depth of 39.4mmag. Oshagh et al. (2013) find that transit-timing variations of up to 200s can be produced when a spot is completely dark and has the size of the largest spot on the Sun. With multiple observation of the same transit we would be sensitive to such effects. However, our rather sparse dataset in particular lacking multiple transits of the same star, does not warrant such investigations. Where the information is available, the online tool modifies the predicted errors account for the activity of target star or gives a warning. Oshagh et al. (2013) assert that even a sophisticated fitting procedure for the depth and time of each individual transit will not produce accurate results for active stars. In practice the prediction will always be limited by the quality of the input assumptions.

The key feature of the online tool is not the analysis of the transit event per se but its ability to offer a useful prediction of what the transit might look like before attempting to capture the transit. This helps the observer to properly plan their observations and for example to adjust the exposure time to see if a relatively short or long exposure time is better when considering transit duration, image cadence and other photometric uncertainties. This functionality of the tool is illustrated in Figure 4 that provides a breakdown of the error with the CKT telescope over a range of catalogue magnitudes, using appropriate exposure times for the magnitudes and where more than one exposure time is available taking the average value. The change in appearance of the percentage error after 12th magnitude is caused by fixing the exposure time at 300s corresponding to the approximate length of time that a small telescope might guide for without intervention. 

In Figure 5, we show how the online tool maybe used to predict system performance in mmags as a route to adjusting observations in order that a particular precision is obtained. Figure 5(a) illustrates the predicted performance, without a contribution to the error from the reference star, over a range of catalogue magnitudes and exposure times. For clarity, the performance has been set to zero where there is no significant predicted performance. Figure 5(b) illustrates the corresponding predicted performance where the precision is better than 3mmag to illustrate the operating region where 3mmag precision is required to adequately capture an exoplanet transit. Alternative plots can be generated with different precision requirements.

\section{Discussion}\label{Discussion}

We have shown that despite using data for a relatively poor site for weather, by following a set of procedures one can derive the predicted precision of an observation that matches the observed precision, for a large number of images taken with a wide range of catalogue magnitudes, exposure times, air masses and observing conditions. The formulation has been successfully used to assist in the satisfactory capture of an eclipsing binary with multiple telescopes. This allows merging results from different telescopes with known precision for each measurement. The formulation has been implemented into an online tool that can be adapted for different telescope and camera configurations to readily predict expected precisions for a wide variety of user- specified assumptions. 

With the ability to predict observational precision for a specific setup, the online tool has been successfully used to assist in the satisfactory capture of several exoplanet transits. It can effectively identify those exoplanet transits that can be adequately captured by a particular telescope and camera for a wide variety of practical limitations or most importantly those that are unsuitable for observing. In particular, the tool can optimise the exposure time to give a satisfactory cadence without exceeding any other practical limits, e.g., the linearity limit of the camera, air mass, off-transit measurements and duration for a proposed set of observations.

The large number of transit events of high and low quality being produced by large scale surveys, e.g. the Transiting Exoplanet Survey Satellite (TESS) means that it is highly desirable to increase the number of observatories able to make routine transit observations.  Similarly there is a variety of exoplanet transit follow-up that requires multi-site observations, e.g. Baluev et al (2015).  

Non-professional observatories have provided many important contributions to the discovery and follow-up of exoplanet transits such as in Burdanov et al (2018) and Tan (2018) though it is notable that no first author publications appear to have resulted. This work joins a variety of resources both online, e.g. Planethunters (2018), Jensen (2013) and offline, e.g., Gary (2014) and Conti (2018), and is intended to provide a practical route to support the many observatories which do have suitable equipment but lack the variety of knowledge necessary to plan and execute transit observations.

\section*{Acknowledgements}

We are grateful for the reviewer, for their many helpful and insightful comments on the original draft of the manuscript. The Exoplanet transit tool of the Czech Astronomical Society has been invaluable in developing our online resource. This research has made extensive use of the SIMBAD database, operated at CDS, Strasbourg, France and the NASA Exoplanet Archive and the Exoplanet Follow-up Observation Program websites, which are operated by the California Institute of Technology, under contract with the National Aeronautics and Space Administration under the Exoplanet Exploration Program. LR is supported by a Brian May scholarship. HJ acknowledges support from the UK Science and Technology Facilities Council [ST/M001008/1 and ST/R006598/1].

\section*{References}
Baluev, R.V., Sokov, E.N., Shaidulin, V.S., et al., 2015, MNRAS 450, 3101

Batalha, N.M., Borucki, W.J., Koch, D.G., et al., 2010, ApJL, 713, L109

Balona L.A., Janes K., Upgren A.R., 1995, in: New Developments in Array Technology and Applications, Philip A.G.D. ed. p. 187

Bayfordbury-Observatory, 2019, Bayfordbury AllSky Camera, Bayfordbury Observatory, University of Hertfordshire [http://star.herts.ac.uk/allsky/]

Beck, P.J., 2018, Precision Photometry at the University of Hertfordshire's Bayfordbury Observatory, MSc Thesis, University of Hertfordshire [http://hdl.handle.net/2299/19912]

Burdanov, A., Benni, P., Sokov, E., et al., 2018, PASP, 130, 074401

Carrington R.C., 1991, Gallic War Book 5 \textendash  BCP Latin Texts. Volume Editor Carrington R.C., Bloomsbury

Castellano T., Laughlin G., Terry R.S., Kaufman M., Hubbert S., Schelbert G.M., Bohler D., Rhodes R., 2004, JAAVSO, 33, 1

Cavazzani, S., Ortolani, S., Zitelli, V., \& Maruccia, Y., 2011, MNRAS, 411, 1271

Cavazzani, S., Zitelli, V., Ortolani, S., 2015, MNRAS, 452, 2185

Collins, K.A., Kielkopf J.F., Stassun, K.G., Hessman, F.V., 2017, AJ, 153, 77

Croll, B., Rappaport, S., Levine, A.M., 2015, MNRAS, 449, 1408

Charbonneau, D., Brown, T.M., Latham, D.W., \& Mayor, M., 2000, ApJL, 529, L45

Conti, D.M., 2018, A Practical Guide to Exoplanet Observing [www.astrodennis.com]

Deeg, H.J., 2013, EPJ Web of Conferences 47, 10001

Dravins D., Lindegren L., Mezey E., Young A.T., 1998, PASP, 110, 610

ExoFOP, 2019, [https://exofop.ipac.caltech.edu/tess/]

Gary, B.L., 2014, Exoplanet Observing for Amateurs. CreateSpace Independent Publishing Platform

Gilliland R.L., Brown T.M., 1992, PASP, 104, 582

Hartman J.D., Stanek K.Z., Gaudi B.S., Holman M.J., McLeod B.A., 2005, AJ, 130, 2241

Henry, G.W., Marcy, G.W., Butler, R.P., Vogt, S.S., 2000, ApJ, 529, L41

Jensen, E., 2013, Astrophysics Source Code Library, 1306.007 [https://github.com/elnjensen/Tapir]

Maciejewski, G., Raetz, St., Nettelmann, N., Seeliger, M., Adam, C., Nowak, G., Neuh{\"a}user, R., 2011, A\&A, 535, 7

NASA Exoplanet archive, 2019, [https://exoplanetarchive.ipac.caltech.edu/]

Oshagh, M., Santos, N.C., Boisse, I., Boue, G., Montalto, M., Dumusque, X., Haghighpour, N., 2013, A\&A, 556, 19

Planethunters, 2018, Online exoplanet database, in Planet Hunters,  [https://www.planethunters.org/]

Poddany S., Brat L., Pejcha O., 2010, New Astronomy 15, 297 [http://var2.astro.cz/ETD/]

Pollacco, D.L., Skillen, I., Collier Cameron, A., et al., 2006, PASP, 118, 1407

Ricker, G.R., Latham, D.W., Vanderspek, R.K., et al., 2009, BAAS, 41, 193

Ryan P., Sandler D., 1998, PASP, 110, 1235

Southworth J., et al., 2009, MNRAS, 396, 1023

Tan, T.G., 2018, The Perth Exoplanet Survey Telescope, http://pestobservatory.com/discoveries/

Vanhuysse, M., Cales, J.-P., Technologie, C., Santerne, A., Moutou, C., 2011, Exoplanets Photometry with Remote Observatory, EPSC-DPS Joint Meeting, 1456

Young A.T., Dukes R.J., Jr., Adelman C.J., 1992, Automated Telescopes for Photometry and Imaging, p. 73

\end{document}